\documentclass[useAMS,usenatbib]{mnras}

\usepackage{graphics}
\usepackage{amssymb}
\usepackage{amsfonts}
\usepackage{wasysym}
\usepackage{epsfig}
\usepackage{breakurl}

\usepackage{longtable}
\usepackage{times}
\usepackage{mathptmx}
\usepackage[usenames,dvipsnames]{color}
\usepackage{url}

\newcommand{\roa}{{$\rho$\,OphA}}

\newcommand{\roc}{{$\rho$\,OphC}}

\newcommand\bz{$\langle B_{\mathrm z}\rangle$}

\title[The auroral radio emission of \roc]
{The auroral radio emission of the magnetic B-type star \roc}

\author[P. Leto et al.]
{P. Leto,$^{1}$ \thanks{E-mail: paolo.leto@inaf.it}
C. Trigilio,$^{1}$
C. S. Buemi,$^{1}$
F. Leone,$^{2}$
I. Pillitteri,$^{3}$
L. Fossati,$^{4}$
F. Cavallaro,$^{1}$
\newauthor L. M. Oskinova,$^{5,6}$
R. Ignace,$^{7}$
J. Krti\v{c}ka,$^{8}$
G. Umana,$^{1}$
G. Catanzaro,$^{1}$
A. Ingallinera,$^{1}$
\newauthor F. Bufano,$^{1}$
S. Riggi,$^{1}$
L. Cerrigone,$^{9}$
S. Loru,$^{1}$
F. Schillir\'{o},$^{1}$
C. Agliozzo,$^{10}$
\newauthor N. M. Phillips,$^{10}$ 
M. Giarrusso,$^{11}$
J. Robrade$^{12}$
\\
$^{1}$INAF - Osservatorio Astrofisico di Catania, Via S. Sofia 78, 95123 Catania, Italy\\
$^2$Universit\'{a} di Catania, Dipartimento di Fisica e Astronomia, Sezione Astrofisica, Via S. Sofia 78, I-95123 Catania, Italy\\
$^{3}$INAF - Osservatorio Astronomico di Palermo, Piazza del Parlamento 1, 90134 Palermo, Italy\\
$^4$Space Research Institute, Austrian Academy of Sciences, Schmiedlstrasse 6, A-8042 Graz, Austria\\
$^5$Institute for Physics and Astronomy, University Potsdam, 14476 Potsdam, Germany\\
$^6$Kazan Federal University, Kremlevskaya Str 18, Kazan, Russia\\
$^7$Department of Physics \& Astronomy, East Tennessee State University, Johnson City, TN 37614, USA\\
$^8$Department of Theoretical Physics and Astrophysics, Masaryk University, Kotl\'{a}\v{r}sk\'{a} 2, CZ-611 37 Brno, Czech Republic\\
$^{9}$Joint ALMA Observatory, Alonso de C\'{o}rdova 3107, Vitacura, Santiago, Chile\\
$^{10}$European Southern Observatory, Karl-Schwarzschild-Strasse 2, Garching bei M\"{u}nchen, 85748, Germany\\
$^{11}$INFN, Laboratori Nazionali del Sud, Via S. Sofia 62, I-95123 Catania, Italy\\
$^{12}$Hamburger Sternwarte, University of Hamburg, Gojenbergsweg 112, D-21029 Hamburg, Germany
}
\begin{document}

\date{}

\pagerange{\pageref{firstpage}--\pageref{lastpage}} \pubyear{}

\maketitle

\label{firstpage}


\begin{abstract}
The non-thermal radio emission of main-sequence early-type stars is a signature of stellar magnetism.
We present multi-wavelength (1.6--16.7 GHz) ATCA measurements of 
the early-type magnetic star \roc, which is a flat-spectrum non-thermal radio source.
The \roc\ radio emission is partially circularly polarized with
a steep spectral dependence:
the fraction of polarized emission is about $60\%$
at the lowest frequency sub-band (1.6 GHz) while is undetected at 16.7 GHz.
This is clear evidence of coherent Auroral Radio Emission (ARE) from the \roc\ magnetosphere.
Interestingly, the detection of the \roc's ARE is not related to a peculiar rotational phase.
This is a consequence of the stellar geometry,
which makes the strongly anisotropic radiation beam 
of the amplified radiation always pointed towards Earth.
The circular polarization sign evidences mainly amplification of the ordinary mode of the electromagnetic wave,
consistent with a maser amplification occurring within dense regions.
This is indirect evidence of the plasma evaporation from the polar caps, 
a phenomenon responsible for the thermal X-ray aurorae.
\roc\ is not the first early-type magnetic star showing the O-mode dominated ARE
but is the first star with the ARE always on view.
\end{abstract}


%
%
%
\section{Introduction}

Magnetic early-type (OBA) stars show a typical
rotational variability observed in many bands of the electromagnetic spectrum,
from the UV \citep{shore81} to the radio domain \citep{leone91}.
Such rotational modulation is well explained in the framework of the Oblique Rotator Model (ORM),
where the overall stellar magnetic field topology is described by an almost dipolar field (at kG level)
with the magnetic axis not-aligned with the stellar rotation axis \citep{babcock49,stibbs50}.

About 25\% of the known magnetic early-type stars are
non-thermal radio sources
\citep{drake_etal87,linsky_etal92,leone_etal94}.
In many cases such kind of stars are also X-ray sources \citep{naze_etal14}.
The main observed features, from X-ray to radio wavelengths, are explained
within a common framework outlined by the magnetically confined 
wind shock (MCWS) model \citep{babel_montmerle97}.
Inside the ``inner magnetosphere'' the magnetic field
traps and channels the ionized matter.
The wind plasma streams from opposites hemispheres are then forced to move along the dipolar magnetic field lines.
At the magnetic equator the streams collide and shock, radiating thermal X-rays. 
Far from the star, the magnetic field is no longer able to confine the plasma.
In the resulting current sheets, electrons can be accelerated up to relativistic energies.
These non-thermal electrons
move within the ``middle magnetosphere'' and radiate through incoherent 
gyro-synchrotron mechanism.
The above scenario has been confirmed by 
a 3D modeling approach, where
the radiative transfer equation for the gyro-synchrotron emission mechanism has been integrated along 
the line of sight producing the synthetic brightness spatial distribution and total flux
from the co-rotating stellar magnetosphere \citep{trigilio_etal04,leto_etal06}.

Among the class of the early-type magnetic stars,
six are known to behave like radio pulsars 
\citep{trigilio_etal00,das_etal18,das_etal19a,das_etal19b,leto_etal19,leto_etal20}.
These six sources show low frequency highly polarized pulses,
explained as the stellar analog of the coherent Auroral Radio Emission (ARE)
observed on the magnetized planets of the solar system \citep{zarka98}.
The ARE arises from magnetospheric regions above the auroral ovals, located at the planetary surface,
and is observed from infrared to X-ray bands  \citep{badman_etal15}.

The amplification mechanism of the ARE is the Electron Cyclotron Maser (ECM) powered by an unstable energy distribution, 
which can be developed by the electrons moving within 
a magnetospheric cavity
\citep{wu_lee79,melrose_dulk82}.
The ECM emission mechanism amplifies the radiation
at frequencies close to the first few harmonics of the local gyro-frequency
$\nu_{\mathrm B} = 2.8 \times 10^{-3} B/{\mathrm G}$ GHz,
hence the frequency of the ARE is directly related to the local magnetic field strength. 
The ECM mainly amplifies just one of the two magneto-ionic modes (each one with opposite circular polarization direction), 
resulting in the high polarization degree of the ARE.
The mainly amplified magneto-ionic mode depends on the density of local thermal plasma 
\citep{sharma_vlahos84,melrose_etal84}.

The ECM theory predicts that the amplified radiation is constrained within a strongly anisotropic beam 
shaped like a thin hollow cone of large half-aperture ($\approx 90^{\circ}$) and
centered to the magnetic field line. The ECM ray path 
is therefore oriented almost perpendicularly with respect to the local magnetic field vector, 
making the ARE observable only when the magnetic field vectors of the auroral source regions are almost perpendicular to the line of sight.
The ARE from the early-type magnetic stars is produced in thin dipole shaped magnetic cavity, the middle-magnetosphere.
The elementary ECM sources aligned along the direction 
that maximizes the path inside the thin auroral cavity 
coherently contribute to produce the detectable ARE.
Hence, in case of ECM amplified within a thin magnetic shell, the overall ARE beam is mainly oriented
along the plane tangent to the auroral cavity wall, like the case of the Earth's auroral kilometric radiation \citep{louarn_lequeau96}.
Interestingly, in some early-type magnetic stars
the signature of the auroral emission 
was discovered in the X-ray spectra
\citep{leto_etal17,leto_etal18,leto_etal20,robrade_etal18},
and in the case of \roa\ also in its X-ray light curve \citep{pillitteri_etal17}.

In this paper we present new multi-wavelength radio measurements of 
\roc\ (HD\,147932), a magnetic \citep{alecian_etal14} B5V~type star that has also exhibits X-ray pulses \citep{pillitteri_etal16},
like the case of \roa\ \citep{pillitteri_etal17}.
{\roc\ exhibits coherent and incoherent radio emission,
whose study allowed us to characterize its magnetosphere.
The radio and X-ray combined results
allowed us to better understand the plasma processes occurring within the
stellar magnetosphere.}


\section{Radio measurements}
\label{sec_atca}
New L, C, X, and U-bands radio observations (centered at 2.1, 5.5, 9, and 16.7 GHz) of \roc\ were
performed on 2019 March by the Australian Telescope Compact Array\footnote{The Australia Telescope Compact Array is part of the Australia Telescope National Facility which is funded by the Australian Government for operation as a National Facility managed by CSIRO} 
(ATCA).
The data were edited and calibrated using the standard {\sc{miriad}} software package.
The backend system has a bandwidth 2 GHz wide, for each band,
which allows us to better investigate the spectral
dependence of the \roc's radio emission 
by analyzing the 1 GHz wide sub-bands.

\roc\ was observed by ATCA during three epochs, about 10 hrs per epoch,
when the interferometer was pointing towards \roa\ (data already published, see \citealp{leto_etal20} for further technical details).
Hence \roc\ was observed $\approx 2.5\arcmin$ away from the phase tracking center.
To obtain reliable flux density measurements, we performed
the primary beam correction using the task {\sc{linmos}}.
The correction to apply increases with the observing frequency,
as a consequence the noise background level also increases.
The U-band measurements were performed using the 15mm receiver 
that acquires 2 simultaneous bands at 16.7 (U-band) and 21.2 GHz (K-band).
Unfortunately \roc\ 
is close to the primary beam first null at the K-band,
hence we were not able to retrieve the highest frequency band of the 15mm receiver.

The ATCA is a linear interferometer that precludes making maps with short integration time
(source snapshots), therefore, we retrieved the time dependent flux density measurements of \roc, both for the Stokes\,$I$ and $V$,
by means of the discrete Fourier transform of the complex visibilities at the source position
\citep{trigilio_etal08,trigilio_etal18,leto_etal20}.
No sources are present around the field centered on \roc\
(the closest radio source is \roa) enabling us to rely on the time resolved measurements.

\section{The radio emission of \roc}
\label{sec_radio}
To search for possible rotational modulation of the radio emission of \roc,
we phase folded the time resolved multi-wavelength ATCA measurements.
The photometric time series provided by the Kepler's K2 mission
clearly evidence the rotation period of  $P=0.8639$ days
\citep{rebull_etal18}.
The phase folded ATCA observations uniformly sample the rotational period.
The zero phase is the start of the ATCA observations: ${\mathrm{HJD}}=2458555.073$. 

\begin{figure}
\resizebox{\hsize}{!}{\includegraphics{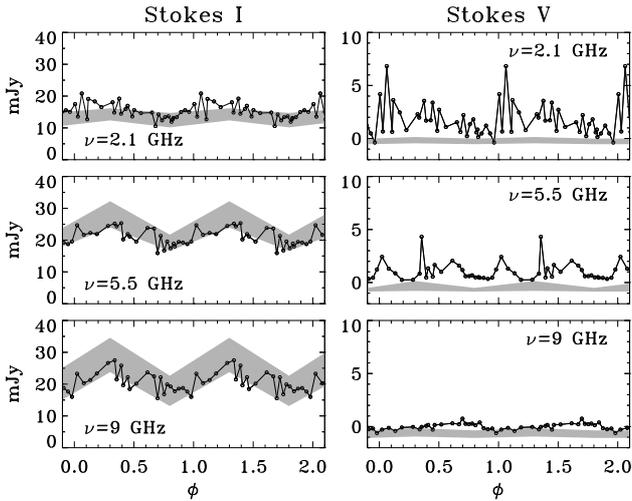}}
\caption{Left panels:  measured multi-wavelengths phase-folded light curves
of the total intensity (Stokes\,$I$) of \roc\
(integration time $\approx 15$ minutes).
Right panels: light curves of the circularly polarized emission (Stokes\,$V$).
The gray regions show the theoretical incoherent gyro-synchrotron emission (both for the Stokes\,$I$ and $V$) 
simulated using the stellar parameters of \roc.
}
\label{l_curve}
\end{figure}

The wide band radio emission of \roc\ is roughly modulated by the stellar rotation, see Fig.~\ref{l_curve},
similar to other magnetic early-type stars well studied at the radio band
\citep{trigilio_etal04,leto_etal06,leto_etal12,leto_etal17,leto_etal18,leto_etal20,bailey_etal12}.
We also observed large variation between measurements performed at similar phases,
this suggests that some other kind of variability mechanism overlaps with the rotational modulation. 

\begin{figure}
\resizebox{\hsize}{!}{\includegraphics{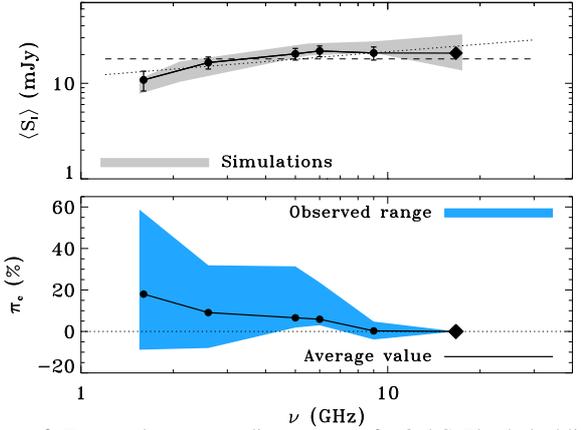}}
\caption{Top panel: average radio spectrum of \roc.
The dashed-line is a perfectly flat spectrum, the dotted-line is the fit performed using a power-law.
The gray area is the envelop of the simulated spectra performed using different model parameters.
Bottom panel: spectrum of the fraction of the circularly polarized emission.
The thick solid line show $\pi_{\mathrm c}$ averaged over the whole rotation period.
The blue area locates the $\pi_{\mathrm c}$ extrema measured at the different observing frequencies.
The diamond symbol in both panels refers to the U-band measurement.
}
\label{fig_spec}
\end{figure}

To better analyze the radio spectrum 
the ATCA measurements at the L and C bands 
are splitted each in two sub-bands 1 GHz wide
(centered at $\nu=1.6$ and 2.6 GHz for the L band; at $\nu=5$ and 6 GHz for the C band).
The corresponding total intensities, 
average of the time resolved radio measurements of all the epochs,
are shown in the top panel of Fig.~\ref{fig_spec}.
The flux of \roc\ at the U-band ($\nu=16.7$ GHz) is $S_{\mathrm I}=20.7 \pm 0.2$ mJy,
this value was measured from the U-band radio map performed collecting all the observing epochs,
the corresponding $S_{\mathrm V}$ is below the detection threshold.

The average radio spectrum of \roc\ is quite flat with 
a positive spectral index (Fig.~\ref{fig_spec} top panel). 
The average total flux density (frequency range 1.6--16.7 GHz)
is $\langle S_{\mathrm I}\rangle \approx 18 $ mJy.
Then, at the distance $d=134$ pc \citep{gaia_dr2}, the spectral radio luminosity 
of \roc\  is $L_{\nu,\,{\mathrm {rad}}}\approx4\times10^{17}$ erg s$^{-1}$ Hz$^{-1}$.
Luminosity level and flat spectrum are absolutely normal for such kind of magnetic stars
{\citep{leto_etal17,leto_etal18,leto_etal20}.}

The  \roc\ low frequency emission is circularly polarized,
with the polarization level that steeply decreases as the observing radio frequency increases.
Figure~\ref{fig_spec} (bottom panel) shows the 
spectral dependence of the measured range of the circular polarization fraction 
$\pi_{\mathrm c}$ ($S_{\mathrm V} / S_{\mathrm I}$).
The sub-band centered at 1.6 GHz evidence the highest level of circularly polarized emission,
in particular the highest measured level  
was $\pi_{\mathrm c} \approx 60\%$. 
The Right hand Circularly Polarized (RCP) emission is dominant (Stokes\,$V>0$).
Such high level of circular polarization of the low frequency emission 
has already been observed
in six other early-type magnetic stars.
Acting as radio lighthouses, these stars present
highly polarized radio pulses.
Such behavior has been explained as stellar auroral radio emission \citep{trigilio_etal11}
powered by the ECM coherent emission mechanism.
The observed high circular polarization level suggests
that the magnetosphere of \roc\ is also a source of coherent auroral radio emission.
However, unlike the other magnetic stars,
the detection of the ARE from \roc\ does not seem to be related to a particular stellar rotational phase.
Looking at the right panels of Fig.~\ref{l_curve}, it is evident that  
the \roc\ polarized emission is not constrained to specific phase ranges, 
as opposed to the coherent pulses of the other six early-type magnetic stars sources of stellar ARE,  
that are observed at well defined phases.

\section{Radio emission Modeling}
\label{sec_magsph}
\subsection{Coherent auroral emission}
\label{coe_simul}
To further investigate the nature of the low frequency emission mechanism operating in \roc,
we performed an extensive set of simulations
of the auroral radio emission visibility from a dipole shaped stellar magnetosphere
(see \citealp{leto_etal16} for details).
The ARE beam pattern is strongly anisotropic, this is defined by the
hollow-cone half aperture ($\theta_{\mathrm {B}}$) and thickness ($\Delta \theta$),
and by the opening {angle ($\delta$)} of the radiation diagram centered on the plane 
tangent to the auroral cavity.

X-ray variability of \roc\ closely resembles the case of \roa.
In case of \roa\ this variability was recognized as signature of the X-ray counterpart of the ARE.
The signature of X-ray auroral emission was also attributed to
the hottest thermal component ($>40$ MK) needed to fit the X-ray spectrum of \roa. 
Also in case of \roc\, the X-ray spectral model fitting suggests the presence of gas
at temperature $\approx 60$ MK (taken from Table~D.1 of \citealp{pillitteri_etal16}).
Similarly to \roa, which has its X-ray emission mainly sustained by auroral emission \citep{leto_etal20},
the high temperature plasma component in the \roc's X-ray spectrum
is likely produced by those relativistic electrons responsible of all
the features observed at radio wavelengths. 
The above considerations can be used to constraint the ARE beam parameters.
According to the ECM theory
the angles $\theta_{\mathrm  {B}}$ and $\Delta\theta$ are related to the radiating electrons speed ($v$):
$\theta= \arccos v/c$;  $\Delta\theta \approx v/c$ ($c$ is the speed of light).
For electrons with energy $\approx 5$ keV ($T\approx 60$ MK), 
it follows: $\theta_{\mathrm {B}} \approx 82^{\circ}$; $\Delta\theta \approx 8^{\circ}$.

\begin{figure}
\resizebox{\hsize}{!}{\includegraphics{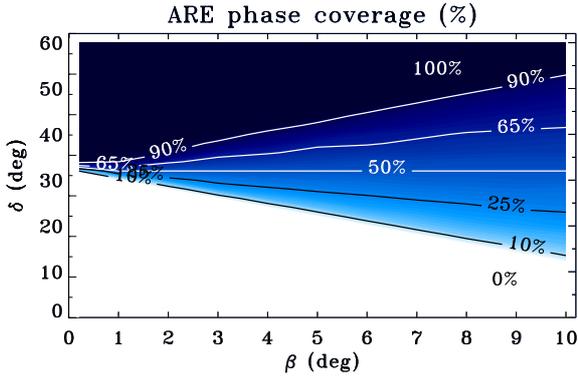}}
\caption{Phase coverage of the ARE from \roc\ simulated varying the parameters $\beta$ {and $\delta$}.
The dark blue region locates the parameter combinations that make the ARE always detectable.
The parameter combinations in the white area precludes the ARE detection.}
\label{fig_sim_are}
\end{figure}

Because of the high ARE directivity,
the ORM geometry is crucial for the auroral emission detection. 
We derived the inclination of the stellar rotation axis
(angle $i$) of \roc\ using the relation $v\sin i = 2 \pi R_{\ast} \sin i /P_{\mathrm{rot}}$.
Adopting the projected rotational velocity $v\sin i=186$ km\,s$^{-1}$ \citep{jilinski_etal06} 
and the stellar radius $R_{\ast}=3.3$ R$_{\odot}$ \citep{gaia_dr2},
it follows $i \approx 74^{\circ}$.
\roc\ is a magnetic star, two longitudinal average magnetic field measurements (\bz), separated by one day, were reported by \citet{alecian_etal14}.
The \bz\ values are both negative and with comparable strength ($\approx -1$ kG).
The magnetic
field axis orientation (angle $\beta$ between magnetic and rotation axes)
has been varied in the range $0^{\circ} < \beta < 10^{\circ}$.
Larger values of $\beta$ produce large \bz\ rotational modulation
unable to reproduce the almost constant \bz\ measurements.
The corresponding polar magnetic field strength ($B_{\mathrm p}$) lies in the range 12--15.8 kG. 
The other free parameter is the tangent plane opening {angle $\delta$}, 
that was linearly increased from $5^{\circ}$ up to $60^{\circ}$ (step $1^{\circ}$).
Results of our simulations are shown in Fig.~\ref{fig_sim_are}.
We found that, once fixed $\beta$, there are values of {$\delta$ able} to produce
ARE visible for a large fraction of the stellar rotation.
The sources of detectable ECM are mainly located in the southern hemisphere, but visible sources of ECM located in the northern hemisphere exist,
this reduces the polarization level of the measured ARE.

\subsection{Incoherent gyro-synchrotron emission}
\label{gyro_simul}
Once constrained the ORM geometry of \roc, 
we simulated its incoherent gyro-synchrotron emission
adopting the central values derived in Sec.~\ref{coe_simul}, 
$\beta=5^{\circ}$ and $B_{\mathrm p}=13.2$ kG.
Our 3D model is able to simulate the brightness spatial distribution and the total flux produced
by a population of non-thermal electrons trapped in the middle magnetosphere. 
Following the MCWS scenario, the inner magnetosphere is filled by thermal electrons with a
temperature that increases linearly outward,
starting from the stellar temperature
$T_\mathrm{eff}= 17 \pm 1$ kK \citep{alecian_etal14}.
Their number density is instead inversely related to the radial distance.
We performed simulations varying the electrons density at the stellar surface ($n_{0}$) 
in the range $10^7$--$10^{10}$ cm$^{-3}$.
The non-thermal electrons are power-law energy distributed ($E^{-p}$) with
{the index $p$} in the range 2--4.

The equatorial extension of the middle magnetosphere
is given by the Alfv\'{e}n radius ($R_{\mathrm{A}}$),
that locates the distance at the magnetic equatorial plane
where the plasma pressure equates the magnetic energy density.
Outside the Alfv\'{e}n radius the magnetic field strength will be unable to force the plasma to co-rotate with the star,
this is the acceleration site of the non-thermal electrons.
We varied $R_{\mathrm{A}}$ in the range 10--30 R$_{\ast}$ (step 5 R$_{\ast}$).
The linear size of the non-thermal electrons acceleration region,
corresponding to the equatorial thickness ($l$) of the middle-magnetosphere,
and the relativistic electrons density ($n_{\mathrm r}$) are degenerate parameters.
Hence, we varied the column density $n_{\mathrm r} \times l$ of the non-thermal electrons 
located at $R_{\mathrm{A}}$.

{Among the full range of the explored model parameters, we found some sets of parameters}
able to roughly reproduce
the observed rotational modulation and spectrum for the total intensity radio emission.
The simulated light curves (both for the Stokes\,$I$ and $V$) and the corresponding average spectra
are shown in Figs.~\ref{l_curve} and~\ref{fig_spec} (gray regions).
Model solutions are obtained using $R_{\mathrm A} = 15$ R$_{\ast}$ and
$n_{0}=1$--3$\times 10^9$ cm$^{-3}$.
The column density $n_{\mathrm r} \times l$ lies in the
range $10^{14.9}$--$10^{17.6}$ cm$^{-2}$,  {as $p$ increases} from 2 to 4.
This is a consequence of the highest energy electrons decay
when the energy spectrum becomes steeper.

Conversely to the Stokes\,$I$ (left panels of Fig.~\ref{l_curve}), 
the Stokes\,$V$ simulated level of the incoherent gyro-synchrotron emission is far from the measured level of 
the \roc\ polarized emission, mostly at the lower frequencies
(see right panels of Fig.~\ref{l_curve}).
This is further evidence that an additional emission mechanism is in act within the magnetosphere of \roc, 
namely the coherent ARE.

\section{Results}
\label{results}
The \roc's magnetosphere is viewed almost equator on
with the southern magnetic pole always visible.
The simulations of the ARE visibility locate
the sources of detectable coherent emission mainly in the southern magnetic hemisphere. 
Furthermore, the observed circular polarization sense is mainly RCP.
These two evidences are in accordance with the ECM mechanism that amplifies the
ordinary magneto ionic mode within the \roc's magnetosphere,
like observed in other two early-type magnetic stars \citep{leto_etal19,das_etal19b}.

The ECM coherent emission mechanism mainly amplifies the O-mode when 
the condition $\nu_{\mathrm p}/\nu_{\mathrm B} >0.3$--0.35 (with $\nu_{\mathrm p}$ the local plasma frequency)
is satisfied \citep*{sharma_vlahos84,melrose_etal84}.
The plasma frequency is related to the thermal plasma density ($N_{\mathrm e}$)
as $\nu_{\mathrm p} = 9 \times 10^{-6} \sqrt N_{\mathrm e}$ (GHz). 
The above condition can be used to constrain the density of the local thermal plasma.
If the first harmonic of the local gyro-frequency amplified by the ECM mechanism is able to escape,
the ARE of \roc\ at $\nu=1.6$ GHz will arise from magnetospheric regions at height
$\approx 1.8$ R$_{\ast}$ above the magnetic pole.
If we observed the second harmonic of $\nu_{\mathrm B}$, 
that is more likely due to the strong gyro-magnetic absorption effects suffered by the $1^{\mathrm {st}}$ harmonic radiation, 
the ARE emitting region will be located at $\approx 2.6$ R$_{\ast}$.
The corresponding electron density needed to produce O-mode amplification for the second harmonic radiation
will be $N_{\mathrm e} > 9.7 \times 10^{8}$ cm$^{-3}$.
 
\begin{figure}
\resizebox{\hsize}{!}{\includegraphics{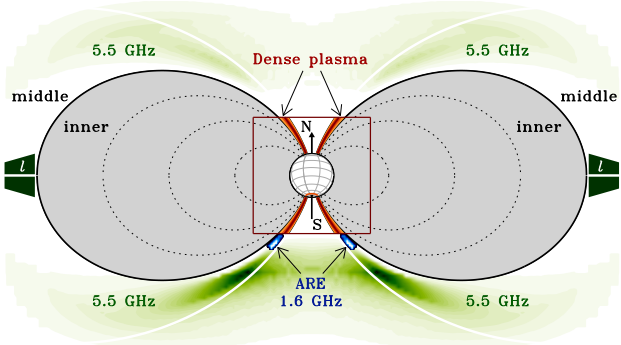}}
\caption{Simulated radio map at $\nu=5.5$ GHz (green) of the \roc\ incoherent gyro-synchrotron emission.
A cartoon showing the 
different regions of the \roc\ magnetosphere was superimposed to the synthetic radio map.
The grey area pictures the inner magnetosphere
where the magnetic field strength is high enough to confine the thermal plasma.
The middle magnetosphere is located outside and is delimited by the white thick lines.
The blue bright spots locate the magnetospheric regions producing the ARE at 1.6 GHz.
The underlying red areas, delimited by the red square,
highlight the middle-magnetosphere regions 
where the thermal plasma density is high enough
to allow the ECM to mainly amplify the O-mode.}
\label{scenario}
\end{figure}

\roc\ is surrounded by a large magnetosphere, 
which emits at the radio regime 
for the incoherent gyro-synchrotron emission mechanism
and the coherent ECM emission mechanism.
The scenario able to summarize the observed radio features of \roc\ is pictured in Fig.~\ref{scenario}.
Following the modeling approach described in Sec.~\ref{gyro_simul} 
we performed the synthetic brightness spatial distribution of the \roc\ incoherent gyro-synchrotron emission
at $\nu=5.5$ GHz.
In Fig.~\ref{scenario}, the simulated radio map
are shown using the green color levels scaled to the highest brightness value.
As additional result of the analysis performed in Sec.~\ref{coe_simul},
the simulated map of the ARE has been performed.
The blue bright regions pictured in Fig.~\ref{scenario} above the south pole locate the source of coherent emission.

The fast electrons, responsible of the gyro-synchrotron emission,
impinging on the surface at the polar caps will be lost.
Hence, the non-thermal electrons reflected by the magnetic mirroring mechanism will be deprived
by the electrons that impact the stellar surface. 
This is a suitable condition for the onset of the unstable loss-cone electrons energy distribution, that
drives the elementary ECM coherent emission mechanism able to power the observed auroral radio emission from \roc.
Meantime, the fast electrons impacting with the stellar surface
might produce plasma evaporation enhancing the thermal electron density 
of the middle-magnetosphere close to the star.
The increased plasma density explains the O-mode maser amplification.

The evaporating hot thermal plasma might be also source of thermal auroral X-rays.
This X-ray source is located within the red box pictured in Fig.~\ref{scenario} .
The emission measure ($EM=N^2_{\mathrm e} V$) of the X-ray source 
is constrained by using the plasma density needed to detect the O-mode of the coherent ARE.
The source volume ($V$) is instead constrained by using 
the model parameters that reproduce the incoherent gyro-synchrotron radio emission of \roc\ (Sec.~\ref{gyro_simul}).
The size of the middle-magnetosphere is $R_{\mathrm A}=15$ R$_{\ast}$,
the equatorial thickness ($l$) of the non-thermal acceleration region was between 0.1--1 $R_{\mathrm A}$.
As previously discussed $l$ cannot  be univocally assigned,
therefore we estimated the $EM$ lower limit in the range $2$--$10 \times 10^{52}$ cm$^{-3}$,
roughly assuming a homogeneous plasma density given by the condition for the 1.6 GHz O-mode amplification.
These estimates are in a very good agreement with the EM derived from the analysis of X-ray observations.
In fact, the hottest thermal plasma component observed in the \roc\ X-ray spectrum
has $EM \approx 9 \times 10^{52}$ cm$^{-3}$, 
derived using the relation $EM = 4 \pi d^2 N \times 10^{14}$, 
where $d$ is the stellar distance and the normalization factor $N=4.2 \times 10^{-4}$ cm$^{-5}$ 
\citep{pillitteri_etal16}.
Comparing the results retrieved by the radio and X-ray observations,
we can constraint the size of the non-thermal acceleration site.
In fact, the measured $EM$ is almost equal to the estimated lower limit retrieved assuming a large middle-magnetosphere thickness. 
Therefore, a limited linear extension for the acceleration site of the non-thermal electrons is more reasonable.

\section{Summary}
\label{discussion}
In this paper we report the detection of auroral radio emission from the early-type magnetic star \roc.
Among this class, \roc\ is the seventh where the stellar ARE has been detected,
but the first where the ARE was visible along the whole stellar rotation period.

The ARE is produced by the coherent ECM mechanism, that is
characterized by a strongly anisotropic emission beam
almost perpendicularly oriented with respect to the magnetic field vector. 
The ARE is detectable when the line of sight is almost perpendicular to the magnetic field axis.
The permanent visibility of the ARE from \roc\ is a consequence of its ORM geometry.
In fact, a dipole magnetic axis almost aligned with the stellar rotation axis,
that in turn has a large inclination angle with the line-of-sight,
is able to explain the observations.

In summary, the theoretical analysis of the ARE visibility condition enables us to constraint the ORM geometry of \roc.
Further, we highlight the magneto-ionic O-mode mainly amplified by the ECM mechanism,
a condition that constrains the density of the local thermal plasma.
Finally, the available observations of \roc\ 
allow us to find out the contribution of the auroral thermal X-ray emission to
the stellar X-ray spectrum.

As a general remark, the present paper confirms 
that combining observations at the extrema of the electromagnetic spectrum
(radio and X-ray) enable ones to efficiently investigate the plasma process 
occurring within the magnetospheres of the early-type magnetic stars.

\section*{Acknowledgments}
We sincerely thank the anonymous referee for his/her very useful and constructive comments and suggestions.
This work has extensively used the NASA's Astrophysics Data System, and the 
SIMBAD database, operated at CDS, Strasbourg, France. 
LMO acknowledges support from
the DLR under grant FKZ\,50\,OR\,1809 and partial support by the Russian
Government Program of Competitive Growth of Kazan Federal University.
JK was supported by grant GA\,\v{C}R 18-05665S.
FL was supported by ``Programma ricerca di ateneo UNICT 2020-22 linea 2''.

\appendix
\section{Data availability}
The data underlying this article are available in the article and in its online supplementary material.



\end{document}